\documentclass[english]{amsart}
\usepackage{amsmath}
\usepackage{amssymb}
\usepackage{graphicx}
\usepackage{amsthm}



\numberwithin{equation}{section}
\numberwithin{figure}{section}
\theoremstyle{plain}

  \theoremstyle{definition}

\makeindex
\makeatother

\usepackage{babel}
  \providecommand{\definitionname}{Definition}
\providecommand{\theoremname}{Theorem}

\begin{document}

\title[Principles Defining Quantum Mechanics]{A Survey of Physical Principles Attempting to Define Quantum Mechanics}

\author{Gary Oas}
\address{Stanford Pre-Collegiate Studies\\
Stanford University\\
Ventura Hall\\
Stanford, CA 94305-4115}
\author[G. Oas, J. A. de Barros]{J. Acacio de Barros} 
\address{School of Humanities and Liberal Studies\\
San Francisco State University\\
1600 Holloway Ave. \\
San Francisco, CA 94132}

\begin{abstract}
Quantum mechanics, one of the most successful theories in the history of science, 
was created to account for physical systems not describable by
classical physics.  
Though it is  consistent with all experiments conducted thus far, 
many of its core concepts (amplitudes, global phases, etc.) can not be directly accessed 
and its interpretation
is still the subject of intense debate, more than 100 years since it was introduced.  
So, a fundamental question is why this particular mathematical model
is the one that nature chooses, if indeed it is the correct model.
In the past two decades there has been a renewed effort to determine what physical or informational principles define quantum mechanics. In this paper, recent attempts at establishing reasonable physical principles are reviewed and their degree of success is tabulated. An alternative approach using joint quasi-probability distributions is shown to provide a common basis of representing most of the proposed principles. It is argued that having a common representation of the principles can provide intuition and guidance to relate current principles or advance new principles. 
The current state of affairs, along with some alternative views are discussed. 
\end{abstract}

\maketitle

\section{Introduction}

It is often stated that the two best scientific theories describing our universe are quantum mechanics and general relativity. The former arose out of a need to describe discrepant phenomena observed in the early part of the 20th Century, and its development history is not as straightforward as the latter. Relativity is often associated to one person, Einstein, and it stemmed from very basic physical principles, such as the relativity and equivalence principles. Quantum mechanics, on the other hand, involved many of the biggest names in physics and required a quarter of a century for its mathematical structure to be conceived, and seems to have no agreed upon basic principles that define it. 

It is not entirely unfair to say that quantum theory was hobbled together from several ideas introduced to explain certain phenomena (among them, the particle nature of light, the wave nature of matter, discretization of atomic energy levels, spin, etc.). Throughout this phase of early quantum theory, models were being proposed to explain physical phenomena, but they did not rely on any fundamental physical principles (this is not too surprising, as the phenomena seemed to fly in the face of accepted principles of classical physics). At the end of this period and upon later refinements a formal mathematical theory was set in place. To this day no experiment has ever been found deviating from quantum theory (within its realm of applicability). 

Through all its success quantum theory leaves many unanswered questions about the nature of the physical world. One of them is the following: {\em what, if any, physical principles define quantum mechanics?} In contrast to relativity, no physical principle has yet been found that picks out quantum theory as the correct model of our universe. This paper is intended to review and discuss recent attempts to arrive at such principles, where  significant progress has been made in the last two decades.

Someone with a strongly pragmatic or positivistic view may question the pursuit of such principles. After all, quantum mechanics is a really good theory, and, in fact, many well-known physicists challenge the worthiness of such a program. 
The purpose of establishing physical principles is to obtain a deeper understanding of quantum theory.
It is understood by most that a modification of current theories is required to model black holes and the early universe, thus having a deeper understanding of one of the other pillar of physics is critical. Having such principles allow for a more efficient exploration of alternative models.

For instance, physical principles are featured prominently in the recent, heated, debate about the nature of black hole horizons. 
The recent `Firewall' proposal \cite{almheiri_black_2013}
was introduced to alleviate contradictions with the nature of entangled particles near the horizon, to maintain the monogamy of correlations \cite{toner_monogamy_2009-1},
 stemming from quantum theory (and is not a physical principle). 
Opponents of this proposal 
cite the gross violation of the equivalence principle, a well-tested physical principle of general relativity. However, no alternatives (except for exotic, nascent proposals \cite{maldacena_cool_2013}) have been offered. 
Having physical principles of quantum theory could shed light on this debate.

In this survey we concentrate on quantum non-locality, contextuality involving two observers
measuring their subsystems in a spacelike separated manner. When seeking physical principles
in this context, it is common to work in the ``device-independent'' framework \cite{scarani_device-independent_2012}, \cite{brunner_bell_2014}, where experiments involve black boxes with only local inputs (choice of a measurement base) and outputs (measurement outcomes). In this way, any constraints imposed are independent of the underlying theory. Note well, there have been axiomatizations (sometimes stated as principles) before, most notably Hardy's five reasonable axioms \cite{hardy_quantum_2001}.
There have also been many constraints placed on the range of quantum systems that do not arise from physical principles but from mathematical principles 
\cite{landau_empirical_1988}, \cite{uffink_quadratic_2002},\cite{wolfe_quantum_2012}, \cite{dzhafarov_no-forcing_2014}, we do not review these here. However, note that no constraint, outside of having
a representation within quantum theory, can precisely characterize the range of quantum correlations.

The proposed principles surveyed here fall within the non-locality scenario and we note the list is not exhaustive.
Attention is towards
those principles that have played an important role and provide the tightest constraints on supra-quantal correlations.
The standard approaches, often involving conditional probability distributions, are introduced as well
as a novel approach involving extended probabilities. 
One purpose of this discussion is to demonstrate that utilizing extended probabilities, specifically, negative probabilities, can provide not only an efficient method to analyzing 
non-local contextuality scenarios, but also provides a unified underlying approach in which comparison of different principles can be carried out.

We begin this paper by introducing the physical system at the center of these investigations. Whenever possible, throughout this paper the quantum mechanical
description is contrasted with an approach based on extended (negative) probabilities. 
The first principle, no-signaling (NS), is introduced and described in the quantum mechanical and negative probability formalism. Further principles are 
introduced and reviewed, namely communication complexity (NTCC), information causality (IC), macroscopic locality (ML), and local orthogonality (LO). We end the paper with some general conclusions.

%
%

\section{Bipartite systems} 

As discussed in our other contribution to this volume \cite{de_barros_examples_2015}, non-locality is perhaps the most astonishing aspect of quantum mechanics. As such, any defining principle for quantum mechanics should be able to explain not-only non-locality, but why quantum systems are not even more non-local (and yet, consistent with relativity) \cite{popescu_quantum_1994}. Therefore, it should come as no surprise that the research of defining principles for quantum mechanics focus on systems that exhibit non-locality. Consequently, in this paper the primary object of study are bipartite EPR-type systems, where the whole system can be split into two subsystems (see \cite{de_barros_examples_2015} for a somewhat more elementary discussion of such systems). Bipartite systems are the simplest systems which can be non-local, and the goal for this section is to present the main concepts and notations relevant for our later discussions involving bipartite systems.  

\begin{figure}[htbp]

      \centering
   \includegraphics{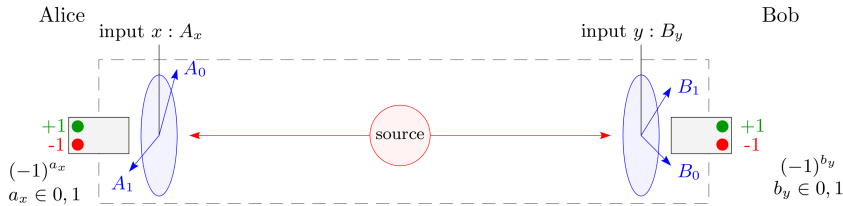} 		
		   \caption{\label{Fig:AliceBob}
 Within the device-independent scenario,
the bipartite two input/output EPR setup consists of a black box where Alice inputs a bit $x\in 0,1$ and receives
output bit $a_x\in 0,1$, while Bob inputs $y\in 0,1$ and receives output $b_y\in 0,1$. The black box is indicated by
the dashed line and the inner workings are ignored and only the statistics of inputs and outputs are analyzed.
The goal is to provide physical principles to constrain results to those achievable by quantum theory. This system
is easily generalizable to more inputs, outputs, and parties, however will not be considered in this discussion.}
   \label{fig:setup}
\end{figure}

For a bipartite system\index{bipartite systems}, let Alice and Bob be two observers located in different places, each receiving one of  two subsystems (see Figure \ref{Fig:AliceBob}).
In a device independent framework\index{device independent framework}, Alice has a choice of input (or experimental setting)  $x \in 0,1$, which results
in an output (or measurement outcome) $\mathbf{A}_x \in \pm 1$. Likewise, Bob inputs $y \in 0,1$ and receives
output $\mathbf{B}_y = \pm 1$. Because the two observers, Alice and Bob, have two inputs and two outcomes, we refer to such setups as 2222 systems.\index{2222 systems} 
Measurement outcomes for a 2222 system are modeled as random variables $\mathbf{A}_x = (-1)^{a_x}, \mathbf{B}_y = (-1)^{b_y}$, where $a_x, b_y \in 0,1$ are output bits. This notation may seem cumbersome, but as it will become clear, it is advantageous to label the random variables by their corresponding bit values $a_x$ and $b_y$.
After many runs of identical states of the bipartite system, the observers, Alice and Bob, can generate statistics to estimate the probabilities $p(a_x,b_y|xy)$ and may attempt to classify the nature of observed correlations. Such statistics are all the (theory-independent) information we gather about those systems. 

If all observed marginal probabilities can be described by a joint probability distribution (jpd) consistent with the standard axioms of probability, we say the system belongs to the  {\em local set} of systems; we denote such set $\mathcal{L}$. 
There are two methods by which local correlations can be established: {\em pre-established strategies} where the
subsystems interact at an earlier time when they were timelike separated\footnote{In computer science parlance, this is called {\em shared randomness}.}; and {\em communication},
where upon measurement information of one subsystem is relayed to the other subsystem and adjustments
are made (by observer or apparatus) to establish the correlation. Since the pairwise measurements under consideration here involve spacelike separated events, communication is not considered as a physically possible method to establish correlations, as it would violate the principles of special relativity.

Local systems can be characterized in different ways, all of them equivalent \cite{fine_hidden_1982}. First, as mentioned above, a system is local if a jpd exists yielding all observable probabilities. Alternatively, locality is equivalent to the system admitting a {\em local hidden variable} model (i.e. the subsystems are independent and can have definite classical states). Finally, such systems are local if they satisfy Bell's inequalities.  

Bell's inequalities, in their original form, are inadequate for experimental systems. A version of Bell's inequalities relevant to our discussion are the CHSH inequalities of the form \cite{clauser_proposed_1969}
\begin{eqnarray}
|S| &=& |\langle \mathbf{A}_0\mathbf{B}_0\rangle  + \langle \mathbf{A}_0\mathbf{B}_1\rangle  + \langle \mathbf{A}_1\mathbf{B}_0\rangle  - \langle \mathbf{A}_1\mathbf{B}_1\rangle | \leq 2. \label{CHSH}
\end{eqnarray}
Because we have $\pm 1$-valued random variables, notice that  $\langle \mathbf{A}_x\mathbf{B}_y\rangle = P(\mathbf{A}_x = \mathbf{B}_y) - P(\mathbf{A}_x\neq \mathbf{B}_y)$, which can be used to go from the joint expectations to probabilities. If a system of random variables $\mathbf{A}_i,\mathbf{B}_j$, $i,j=0,1$, satisfies (\ref{CHSH}), then it belongs to $\mathcal{L}$ (if superluminal communication is prohibited, then the CHSH is a necessary and sufficient condition for membership in $\mathcal{L}$). There are eight CHSH inequalities obtained by permuting the negative sign within (\ref{CHSH}). 

It is instructive to investigate the range of correlated systems in the space of probabilities. 
Pitowski \cite{pitowsky_quantum_1989}
examined the geometry of the CHSH inequalities and introduced the {\em local polytope},\index{local polytope} given simply by the set $\mathcal{L}$, whose
eight facets correspond to one of the eight CHSH inequalities. The 16 vertices that delineates the local polytope $\mathcal{L}$ are given by the deterministic elementary probabilities of the jpd
$p_{a_0a_1b_0b_1}$, where one probability is 1 and the others vanish. In the parlance of quantum information, a system
with probabilities $p(a_x,b_y |xy)$, or jpd $p_{a_0a_1b_0b_1}$, is termed a {\em box}, and the 16 vertices are referred to as {\em deterministic boxes} or $D$-boxes for short.

It is well known that some physical systems described by quantum mechanics can violate Bell inequalities. 
Systems belonging to the {\em quantum set},\index{quantum set} $\mathcal{Q}$, are those that can be written as
\begin{eqnarray}
p(a_x,b_y|xy) &=& tr [ F_x^{a_x}\otimes F_y^{b_y}\rho ],
\end{eqnarray}
where $\rho$ is a quantum state (a Hermitian unit-trace matrix), and $\{ F_x^{a_x},F_y^{b_y} \}$ are positive operator valued measures (POVM)
which satisfy $\sum_{a_x} F_x^{a_x} = \sum_{b_y} F_y^{b_y} = 1, \forall x,y$. 
 
 The quantum set extends beyond $\mathcal{L}$ up to a maximum found by  Tsirelson  $|S| = 2\sqrt 2$ \cite{cirelson_quantum_1980}. 
 Pitowski showed that $\mathcal{Q}$ does not form a polytope
 but a convex set, as well as $\mathcal{L}\subset\mathcal{Q}$. Extremal quantum
 systems at the Tsirelson bound\index{Tsirelson bound} are referred to as {\em Tsirelson boxes,} or {\em $T$-boxes} for short.  
 
One of the central questions in the search for the basic principles that define quantum mechanics pertain to the precise shape of the quantum set's  boundary, $\partial\mathcal{Q}$. In order to fully characterize this question, the first principle we examine, the no-signaling principle, needs to be introduced.

\section{No-Signaling (NS)}\index{no-signaling condition}

In 1994 Popescu and Rohrlich \cite{popescu_quantum_1994} pondered whether special relativity
constrained the total range of correlated systems to those representable by quantum mechanics. The causal structure imposed by special relativity forbids signals to be sent from one place to another with faster-than-light (superluminal) speeds, and attempts to use entangled quantum systems to send signals violating special relativity turned out to be flawed \cite{peres_how_2003}. So, quantum mechanics seemed like a theory that allowed ``spooky" correlations, but not strong enough to violate relativity. Based on this, Popescu and Rohrlich asked wether the following principle defined quantum mechanics:
\begin{quote}
\em Quantum correlations should not permit superluminal signaling.
\end{quote} 
They answered their question in the negative, by showing that there exist correlated systems which do not permit 
signaling and that cannot be modeled by quantum mechanics. In other words, as we will show below, there are no-signaling boxes that have stronger correlation than those permitted by quantum mechanics. 
 
In order to establish communication between two spacelike separated observers using a
correlated system, the marginal probabilities for one observer must differ upon a change in the
input for the other observer. If every observable marginal probability is invariant under a change of
input of the distant observer, no communication is possible. The no-signaling condition in this scenario
is satisfied by those invariant marginal probabilities,
\begin{eqnarray}
p(a_x|xy) &=& \sum_{b=0}^1 p(a_xb_y|xy) = \sum_{b=0}^1 p(a_xb_{y'}|xy'), \;\;\;\;\; \forall a,x,y,y', \nonumber \\
p(b_y|xy) &=& \sum_{a=0}^1 p(a_xb_y|xy) = \sum_{a=0}^1 p(a_{x'}b_{y}|x'y), \;\;\;\;\; \forall b,y,x,x'.\label{NScond}
\end{eqnarray}

 It is clear that the set of non-signaling boxes extends beyond the local and quantum sets, i.e. $\mathcal{L}\subset\mathcal{Q}\subset{NS}$.  The NS condition is not an inequality, and by itself, it does not define a polytope, but is generally used to reduce the dimension of the sets.  If one includes the constraint that observable marginal probabilities must lie within $[0,1]$, then one obtains what is commonly termed the {\em non-signaling polytope}, $\mathcal{NS}$.

Popescu and Rohrlich introduced the maximal non-signaling system, now known as the PR box\index{PR box} (after their names), as
boxes having perfect correlations,
\begin{eqnarray}
\langle A_0B_0\rangle = \langle A_0B_1\rangle = \langle A_1B_0\rangle = +1, \langle A_1B_1\rangle = -1. \label{PRcorr}
\end{eqnarray}
The PR box takes the maximal algebraic value for the CHSH parameter, $S = 4$. There are 8 PR
boxes in the 2222 system, one each lying above a CHSH facet. The PR boxes, along with the 16 $D$-boxes, are vertices forming the non-signaling polytope.\index{non-signaling polytope}

\begin{figure}[htbp]

      \includegraphics[width=\textwidth]{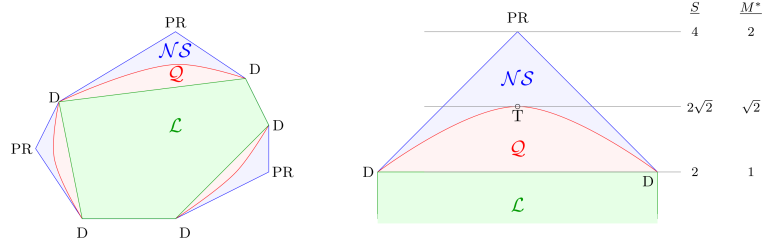} 

   \caption{Within the 16 dimensional space of marginal probabilities, $p(a_x,b_y)$, three subspaces are defined after reducing
the dimensionality to 8 dimensions via the normalization and the no-signaling conditions. The {\em local polytope}, $\mathcal{L}$, are
those systems satisfying the 8 CHSH inequalities, which form the 7-dimensional facets of the polytope. These systems admit
a proper jpd. $\mathcal{L}$ has as vertices the 16 local deterministic boxes, $D$, which have one elementary probability, $p_{a_0a_1b_0b_1}  =1$.
The quantum set $\mathcal{Q}$, consists of those systems that can be represented by states and operators in a Hilbert space. $\mathcal{Q}$
is not a polytope but a convex set. There are eight maximal non-local systems, $T$ boxes, that reach the Tsirelson bound, $S = 2\sqrt 2$ (equivalently,
$M^* = \sqrt 2$). Lastly, the non-signaling polytope, $\mathcal{NS}$, are those which have non-negative marginal probabilities and satisfy
the no-signaling condition. $\mathcal{NS}$ is a polytope formed by the 8 maximally non-local, non-signaling, boxes, labeled $PR$ boxes, and 16 $D$ boxes
as vertices. The sets satisfy $\mathcal{L}\subset\mathcal{Q}\subset\mathcal{NS}$. The figure is a simplified sketch of the 8 dimensional polytope and
in this discussion we will concentrate on one portion, the right figure, of the polytope associated to the CHSH parameter 
$S = \langle A_0B_0\rangle +\langle A_0B_1\rangle +\langle A_1B_0\rangle -\langle A_1B_1\rangle.$}
   \label{fig:NSpolytope}
\end{figure}
 
 To simplify the forthcoming discussion, we limit our focus to just one portion of the non-signaling
 polytope parameterized by $S = \langle A_0B_0\rangle +\langle A_0B_1\rangle +\langle A_1B_0\rangle -\langle A_1B_1\rangle $. 
 The PR box conditional probabilities are compactly represented as 
 \begin{eqnarray}
 P_{PR}(a,b|xy) &=&\frac 12, \;\;\;\;\mbox{if}\;\;\; a\oplus b = xy, \;\; \mbox{0 otherwise}, \label{PRprob}
 \end{eqnarray}
where $\oplus$ signifies addition modulo 2.

\begin{figure}[htbp]
				\centering
			      \includegraphics{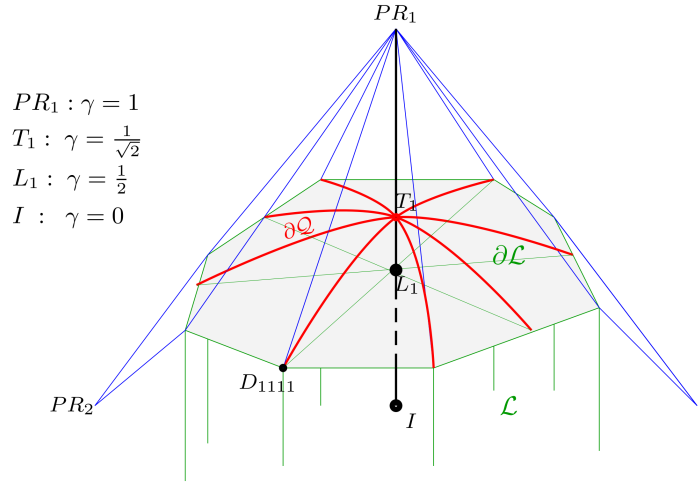} 

    \caption{A more detailed representation of a portion of the $\mathcal{NS}$ polytope highlighting the {\em isotropic} boxes formed
by the linear combination of the {\em noise} box, $I$, (where all correlations are 0) and the $PR^1$ box, $P_\gamma^1 = \gamma PR_1 + (1-\gamma)I$ where
$\gamma = [0,1]$. Figure \ref{fig:NSpolytope}  corresponds to a slice formed by $D_{1111} - L_1 - \overline D_{1111} - PR_1$.}
    \label{fig:PolytopeIsotropic}
 \end{figure}

 There certain trajectories and slices through $\mathcal{NS}$ that will be referred to for the various principles and they are defined here, see
 Figure \ref{fig:PolytopeIsotropic}. The set of {\em isotropic boxes} are those formed by a linear combination of the $PR_1$ box (\ref{PRprob})
 and the {\em noise box}, $I$, which is the system with $P_I(a,b|xy) = \frac 14, \forall a,b,x,y$,
 \begin{equation}
 P_\gamma^1 (a,b|xy) = \gamma PR_1 + (1-\gamma)I. \label{isotropicprob}
 \end{equation}
 The explicit form of the probabilities is given in the following table,
 \begin{equation}
 P_\gamma^1 (a,b|xy) =  \left(\begin{array}{cccc}
\alpha &  \alpha  &  \alpha  & \beta \\
\beta &  \beta  &  \beta  & \alpha \\
\beta &  \beta  &  \beta  & \alpha \\
\alpha &  \alpha  &  \alpha  & \beta \\
 \end{array}\right), \label{P1gamma}
 \end{equation}
where each column is a different input ($(x,y) = (0,0),(0,1),(1,0),(1,1)$), each row indicates a specific outcome $(a,b)$,
 and $\alpha = \frac 14(1+\gamma), \beta = \frac 14 (1-\gamma)$.
 
Along the isotropic trajectory we also identify the box $L_1$, which lies on $\partial\mathcal{L}$, and the {\em Tsirelson box}, $T_1$,
which is the maximally non-local quantum box. Lastly, slices through the polytope will require the introduction of the $PR_2$ box,
which is obtained by swapping the $(0,1)$ and $(1,1)$ columns in (\ref{P1gamma}).

 The failure of no-signaling to define quantum correlations has prompted the fundamental question of what principle, or principles, define the set of quantum correlations.
Due to deterministic correlations and simplicity of the PR box, they will provide an easy first check for violations of proposed principles.

\subsection{Negative Probability}\index{negative probabilities}
No-signaling polytopes include systems whose correlations are too strong to be compatible with a joint probability distribution, as mentioned above. 
Adhering to the device-independent spirit, an approach to exploring those polytopes is to use extended probabilities. There are different ways to extend beyond Kolmogorovian probability. One possible way is to relax the additivity axiom, allowing for sub or super additivity \cite{suppes_existence_1991, de_barros_probabilistic_2010,hartmann_entanglement_2010}, resulting in what is called {\em upper and lower probabilities}. Here we instead keep the additivity axiom but relax the non-negativity requirement, allowing
probabilities of elementary (but unobservable) events to take negative values \cite{oas_exploring_2014,de_barros_negative_2014,de_barros_decision_2014,de_barros_negative_2015,de_barros_quantum_2015}. Only those joint quasi-probability
distributions (jqpd)  admitting nonnegative observable marginal probabilities are considered. 

Relaxing non-negativity gives an extra freedom to the probability function, allowing it to fit the (observable) marginal probabilities that would be inconsistent with a proper jpd. But it also results in an infinite number of possible jqpds yielding observable probabilities. To constrain the number and to provide a measure of deviation from a proper joint distribution (jpd), the L1 norm of the jqpd is minimized \cite{oas_exploring_2014,de_barros_decision_2014,de_barros_unifying_2014}. Explicitly, we minimize $M$, 
\begin{eqnarray}
M^* &=& \min\sum_{a_0a_1b_0b_1} |p_{a_0a_1b_0b_1}|.
\end{eqnarray}
It should be clear that a proper joint distribution, $p_{a_0a_1b_0b_1} = |p_{a_0a_1b_0b_1}|, \forall a_0,a_1,b_0,b_1$ exists if and only if $M^* = 1$ \cite{de_barros_negative_2015}. If there does not exist a proper jpd, yet a jqpd exists yielding the observable marginal probabilities, then $M^*>1$ and the value of the L1 norm can be thought of as a measure of departure from a standard distribution, as mentioned above. 
In fact, it has been shown that for those systems with $M^*>1$ we have $2M^* = S$ \cite{oas_exploring_2014}. Thus, we
can equivalently explore the non-signaling polytope with the CHSH parameter or the L1 norm.

The CHSH inequalities (\ref{CHSH}) can be expressed in terms
of the probabilities of the 16 elementary probabilities, $p_{a_0a_1b_0b_1}$, as $S_{m,n} = 2\sum (-1)^{f_{n,m}}p_{a_0a_1b_0b_1}$
where 
\begin{eqnarray}
f_{m,n} = (a_0\oplus a_1)(b_0\oplus b_1) \oplus a_n\oplus b_m. \label{definef}
\end{eqnarray}
For simplicity, we will
focus on that portion of the non-signaling polytope corresponding to $m=n=0$ and define $f\equiv f_{0,0} = (a_0\oplus a_1)(b_0\oplus b_1) \oplus a_0\oplus b_0$.

It has been shown independently by several authors \cite{abramsky_operational_2014,al-safi_simulating_2013,oas_exploring_2014}
that a necessary and sufficient condition for a system to satisfy the no-signaling condition\index{no-signaling condition} is the existence of a jqpd. In this discussion, we will be limiting to non-signaling systems, 
so that all boxes can be represented by a jqpd\footnote{Though this result seems to imply that negative probability is incapable of examining systems violating
the no-signaling criterion, or better put, those systems having contextuality by direct influence \cite{dzhafarov_contextuality_2014} (or violating marginal selectivity), can be analyzed if one utilizes counterfactual reasoning \cite{de_barros_negative_2015}.
In fact, this is how Feynman\cite{feynman_negative_1987}, and later Scully\cite{scully_feynmans_1994}, analyzed the double slit with negative probability.}.

In exploring the range of non-signaling systems, the isotropic boxes are of central importance. 
Isotropic boxes with minimum L1 norm, $M^*$, are given by the following jqpd, 
\begin{equation}
p_{a_0a_1b_0b_1} = \frac 1{16}(1 + 2\gamma(-1)^{f}), \label{jqpd}
\end{equation}
where $\gamma = 0,\frac 12, \frac 1{\sqrt 2}, 1$ correspond to  the $I, L, T, PR$-boxes respectively
and $f $ is defined in (\ref{definef}) with $m=n=0$.
With this parameterization we have  
\begin{eqnarray}
M^* &=& \frac 12[ 1+2\gamma + |1-2\gamma|].
\end{eqnarray}
The characterization of other trajectories, or subspaces, of the non-signaling polytope can be efficiently analyzed in this manner.

Thus non-local systems can be characterized either by the amount of violation of relevant Bell inequalities, or via the deviation from
a classical system as measured by $M^*$. 

%

\subsection{Characterizing $\mathcal{Q}$}

Before proceeding to further principles an important question must be addressed: Given a set of marginal probabilities, $P(a_x,b_y)$,
how do you determine whether a quantum system can describe it? This is not a trivial question as searching across all possible states, $\rho$,
and all local observables, $E_x, E_y$, is proven hard. 
There have been several works placing necessary constraints on the set of quantum correlations not connected to the search
for physical principles. 
We review those that have been found to be related to some of the principles to be discussed.

Uffink's inequality \cite{uffink_quadratic_2002}
is a constraint satisfied by all quantum correlations, yet is rather weak in limiting supra-quantal systems. It is 
expressed in terms of the second moments $\langle A_iB_j\rangle$ between the random variables
\begin{equation}
(\langle A_0B_0\rangle - \langle A_1B_1\rangle )^2 +  (\langle A_0B_1\rangle + \langle A_1B_0\rangle )^2 \leq 4. \label{uffink}
\end{equation}
The Information Causality principle returns the Uffink inequality along certain trajectories within the non-signaling polytope.

Another strong constraint on supra-quantal systems was independently found by Tsirelson \cite{cirelson_quantum_1980},
Landau \cite{landau_empirical_1988}, 
and later Masanes \cite{masanes_necessary_2003}
(which will be labeled the TLM inequality for short)
and takes the form
\begin{equation}
|\sin^{-1}\langle A_0B_0\rangle + \sin^{-1}\langle A_0B_1\rangle + \sin^{-1}\langle A_1B_0\rangle- \sin^{-1}\langle A_1B_1\rangle|\leq \pi. \label{TLM}
\end{equation}
This inequality is often utilized as an approximation to $\mathcal{Q}$ because of its simplicity and proximity to the quantum set.
This inequality was improved by the work discussed next by substituting the correlation $C_{A_iB_j} = {\langle A_iB_j\rangle - \langle A_i\rangle\langle B_j\rangle \over \sqrt{var(A_i)var(B_j)}}$
for the second moment in (\ref{TLM}). 

 At this time, the best method to characterize systems is via the NPA hierarchy \cite{navascues_bounding_2007}.
Here we give a very simple overview of the hierarchy and refer the reader to their paper for details.
We begin be reiterating the definition of the quantum set, $\mathcal{Q}$.
A system described by a set of marginal probabilities $P(a_x,b_y)$ can be modeled by
quantum mechanics if there exists a normalized state $|\psi\rangle$ in $\mathcal{H}_A\otimes\mathcal{H}_B$,
and a set of Hermitian projection operators $E_x^a, E_y^b$ for Alice and Bob, respectively,  which 
satisfy
\begin{eqnarray}
1) &\;\;& \sum_a E_x^a, = \mathbb{I} \nonumber \\
2) &\;\;& [E_x^a,E_y^b] = 0, \nonumber \\
3) &\;\;& E_x^aE_x^{a'} = 0, \;\;\; \mbox{projectors for distinct outcomes are orthogonal},\nonumber
\end{eqnarray}
and yield the observed marginal probabilities $P(a_x,b_y) = \langle\psi |E_x^a\otimes E_y^b |\psi\rangle$.
Lastly, we include the condition 4) that $\rho$ is a positive, Hermitian, operator.
This method of characterizing quantum correlations (in terms of projectors) is necessary and sufficient if one does not
restrict the dimensionality of the Hilbert space\footnote{This is because the most general type of 
measurement operator, a Positive Operator Value Measure (POVM), can always be cast as a
projective measurement in a higher dimensional Hilbert space.}.

The hierarchy embodies the conditions 1)-4) above by forming a set of projectors $\mathcal{S} = \{S_1, \cdots, S_N\}$ and demanding that
the matrix (or `certificate')
\begin{equation}
\Gamma_{ij} = tr[S_i^\dagger S_j\rho],
\end{equation}
be semi-definite positive. 
The hierarchy forms by considering ever increasing numbers of projectors in the set.
 \begin{eqnarray}
 \mathcal{S}_0 &=& \{ \mathcal{I} \}, \nonumber \\
  \mathcal{S}_1 &=&  \mathcal{S}_0  \bigcup \{ E_x^a \} \bigcup \{ E_y^b \}, \nonumber \\
   \mathcal{S}_2 &=& \mathcal{S}_1  \bigcup \{ E_x^aE_{x'}^{a'} \} \bigcup \{ E_y^bE_{y'}^{b'} \}, \nonumber \\
    \mathcal{S}_3 &=& \cdots 
    \end{eqnarray}
By the properties of the projectors we have $\mathcal{S}_0 \subseteq\mathcal{S}_1   \subseteq \mathcal{S}_2 \cdots$.
Finding the restriction to semidefinite matrices $\Gamma_{ij}$ is a decidable semidefinite programming task (SPD).
The set of systems which admit a semidefinite matrix $\Gamma > 0$ for set $\mathcal{S}_k$ defines the $\mathcal{Q}^k$ level of the NPA
hierarchy approximating the quantum set.
We direct the reader to   \cite{navascues_bounding_2007} for details of determining each level of the hierarchy.

The hierarchy has been proven to be convergent in that  $\mathcal{Q}^k \subseteq \mathcal{Q}^{k-1},  \forall k = 1,..,\infty$. 
In \cite{navascues_convergent_2008} 
it was shown that $\lim_{n\rightarrow\infty} \mathcal{Q}^n = \mathcal{Q}$. Here we will be concerned
with the first level of the hierarchy $\mathcal{Q}^1$ as it approximates $\mathcal{Q}$ rather well and in \cite{navascues_bounding_2007} it was shown that
a necessary and sufficient condition to belong to the $\mathcal{Q}^1$ set of systems is satisfying the TLM inequality (\ref{TLM}) expressed in 
terms of correlations. 

There is one other level that will be of interest and it actually lies between $\mathcal{Q}^1$ and $\mathcal{Q}^2$ and is
 based on the set $\mathcal{S}^{1+AB}$. This set consists of the union of $\mathcal{S}^1$ and the set of operators consisting of all products of two projectors, restricted to one from each party, $E_x^aE_y^b$.
 Thus systems admitting semidefinite certificates for this set of projectors forms the set $\mathcal{Q}^{1+AB}$.

%
%
%
\section{Non-Trivial Communication Complexity (NTCC)}

The next attempt on answering why the class of non-signaling supra-quantal correlations are not
physically obtainable occurred in 2000 with the work of van Dam \cite{van_dam_implausible_2013}. 
He realized that if Alice and Bob share a PR box, then they can compute any Boolean function trivially.
As this seems to be a rather unnatural expectation of physical systems
the following principle, called {\em non-trivial communication complexity}, was proposed:
\begin{quote}
\em For distributed computation tasks the amount of information sent should scale with the size of the task.
\end{quote}

The study of communication complexity in classical and quantum scenarios is a rich field with many results, and here we restrict the discussion to a single task, the {\em inner product game}. Given a distributed set of binary vectors, $\vec v$ for Alice and $\vec w$ for Bob, each with $N$ bits, the goal of this game is to predict the inner product of these two vectors,
\begin{eqnarray}
I &= \vec v\cdot\vec w.
\end{eqnarray}

An easier task is to determine whether $I$ is even or odd.
It is clear that, if Alice and Bob are constrained to only classical resources, Bob must send Alice his $N$ bits in order for Alice to determine whether $I$ is even or odd. Now, consider that instead of only classical resources, Alice and Bob have access to a large number of PR boxes. With access to PR correlations, Alice
can determine the parity of $I$ after receiving only 1 (classical) bit of information from Bob, independent of $N$.
Thus, the original even/odd game becomes trivial with access to PR boxes.

To see this, imagine that Alice and Bob each receive a random binary vector, having components $0$ or $1$, and that they set their input to the $i^{th}$ PR box equal to the $i^{th}$ component of their vector. There are 4 possibilities,
if the components are $(v_i, w_i) = (0,0), (0,1), (1,0)$ then the parity of $I$ does not change (if it was
even/odd it remains even/odd). However, when $(v_i, w_i) = (1,1)$, the parity flips as the inner product increases by 1.
Thus,  the parity of the number of $(1,1)$ input occurrences is equal to the parity of $I$. 
In other words, the PR correlations yield outcomes $(0,0)$ or $(1,1)$ in the first three cases, and the total sum of outcomes does not change the parity (since either 0 or 2 is being added to the sum). But when the inputs are $(1,1)$, the total sum of outputs is incremented by one and the parity flips. Thus, to determine the total parity of outputs, which is equivalent to the parity of $(1,1)$ input events, and which is equivalent to the parity of $I$, Bob only needs to send the parity of the sum total of his
outputs to Alice. 

The following simple example demonstrates how this works. 
\begin{eqnarray}
\vec v = (0,1,1,0,0,1,0), \;\;\;&& \vec w = (1,0,1,1,0,1,0) \nonumber \\
I = \vec v\cdot\vec w = 2 \nonumber 
\end{eqnarray}
In the second column below the $i^{th}$ component of the inner product is given $v_iw_i  = x_iy_i$,
and in the third column are example outcomes satisfying PR correlations.

\vskip 0.1 in 
\begin{tabular}{lclcc}
$i$ \;\;\;& $x_iy_i \;\;$ & $(a_i,b_i)$&& parity \\ \hline
1 & 0 & (1,1)&& 0 \\
2 & 0 & (0,0)&& 0 \\ 
3 & 1 & (0,1)&& 1 \\
4 & 0 & (0,0)&& 1 \\
5 & 1 & (1,0)&& 0 \\
6& 0 & (1,1)&& 0 \\\hline
 &   & $(3,3)$ &&$\downarrow$\\
  &    & {\small Bob sends $3 $ mod$_2 = 1$ to Alice} & \\
  &  &\small  Alice: $\rightarrow 1_A\oplus1_B = 0$: {\bf even!} &&0\end{tabular}
\vskip 0.1 in

In the end, Alice determines the parity of the inner product with probability 1. The most salient point is that this protocol succeeds
regardless of the size of the input vectors $\vec v, \vec w$. NTCC states that such trivial communication tasks should not be physically realizable.

There are theorems stating the isomorphism of the inner product game to a wide range of other algorithmic problems considered in information theory.  Thus, having access to PR boxes makes a large class of computational tasks trivial. (In this survey we have not included the principle No Advantage for Non-Local Computation (NANLC)  for 
there is a close connection between communication and distributed computation tasks.)

The question remains as to how many supra-quantal systems violate this principle and can be considered non-physical. At this date, a portion of the $\mathcal{NS}$ polytope have been eliminated, but not all. Specifically, for isotropic boxes, it has been shown that those having $S > 3.266$ ($M^*>1.633$) admit trivial communication complexity \cite{brunner_nonlocality_2009}. Other portions of the polytope have also been shown to violate NTCC (see figure 4 of \cite{brunner_nonlocality_2009}), however, it has recently been proven that NTCC does not characterize $\mathcal{Q}$ \cite{navascues_almost_2015}.

\section{Information Causality (IC)}

In 2009, Pawloski et. al. \cite{pawlowski_information_2009} considered a communication task akin to others studied in classical computation and applied
it to non-signaling systems. The scenario consists of Alice, who receives a string of $N$ random bits, $\vec\alpha = (\alpha_0, \alpha_1,\cdots , \alpha_{N-1})$, and Bob, who receives a random value $\beta = (0,\cdots, N-1)$ and is tasked to guess Alice's $\alpha_\beta$ bit after Alice has sent $m$ bits of classical information to him. They are allowed  pre-shared correlations, local resources, and access to non-signaling boxes. The degree of success of Bob's guess $g$ is measured by
\begin{eqnarray}
I &\equiv & I(\alpha_K : g|\beta = K),   \label{ICmutualinfo}
\end{eqnarray}
where $I(\alpha_K : g|\beta = K)$ is the Shannon mutual information between $\alpha_K$ and $g$. 
The information causality principle (IC) states that physically allowed theories have $I\leq m$. In other words, 
\begin{quote}
\em One should not expect Bob to gain more information about Alice then Alice has sent classically.
\end{quote}
Note that IC with $m=0$ is simply NS, i.e. without any communication between Alice and Bob, Bob's 
information about Alice's bits should not increase.

\begin{figure}[htbp]

      \includegraphics{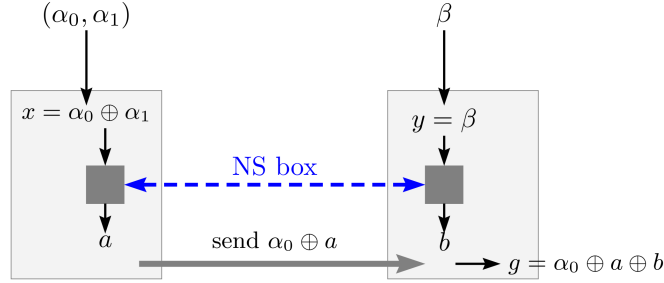}

   \caption{The van Dam protocol for exploring IC. Alice receives two random bits $\alpha_0,\alpha_1$ and Bob is given a random bit $\beta$ with which Bob is tasked
	to guess the $\alpha_\beta$ bit of Alice. They are allowed a single use of a non-signaling box. The protocol consists of Alice inputing $x = \alpha_0\oplus\alpha_1$
	(where $\oplus$ is addition modulo 2). 
	into her side of the box and sending a single bit $\alpha_0\oplus a$ through classical channel to Bob. Meanwhile, Bob inputs his random bit value $\beta$ into
	his side of box and forms the guess by adding (mod 2) his output to the bit received from Alice. It is easy to show that using $PR$ correlations, i.e. $a\oplus b = xy$,
	Bob can always guess the bit $\alpha_\beta$  correctly since $g = \alpha_0\oplus xy$.}
   \label{fig:ICprotocol}
\end{figure}

We examine the simplest case where Alice receives $\vec\alpha = (\alpha_0,\alpha_1), \; \alpha_i\in 0,1$, Bob receives
a single bit $\beta \in 0,1$, 
and Alice sends 1 bit of classical information to Bob. This is illustrated in figure \ref{fig:ICprotocol}.
They are allowed access to a single NS box having inputs $x,y$ and outputs $a,b$ for Alice and Bob respectively.
This box is classified by the set of probabilities $p(a,b|x,y)$ described earlier.

IC has not been examined in full generality and only certain protocols have been explored. A protocol based on communication complexity by van Dam, introduced in the previous section, demonstrates how IC can be violated for a large class of supra-quantal systems. The protocol consists of Alice inputing the bit $x = \alpha_0\oplus\alpha_1$ into her portion of the NS box and Bob inputting $y = \beta$. Alice receives output $a$ and sends the single bit $\alpha_0\oplus a$ to Bob. Bob then constructs his guess as $g = \alpha_0\oplus a\oplus b$. 

Let's examine the probabilities for success of Bob's guess. If Bob inputs $0$ then Bob's guess is correct if $a\oplus b = 0$, i.e. $g = \alpha_0$. Thus, the probability for Bob's success is,
\begin{eqnarray}
P_I \equiv P_{\beta = 0} &=& \frac 12 (p(a\oplus b = 0|0,0)+p(a\oplus b = 0|1,0)). \label{PI}
\end{eqnarray}
If Bob inputs $y=1$ and Alice inputs $x=0$, which implies $\alpha_0=\alpha_1$, then Bob's guess is successful
if, again, $a\oplus b = 0$, since $g = \alpha_0 = \alpha_1$. If Alice inputs $x = 1$, implying $\alpha_0\neq\alpha_1$ or $\alpha_1 = \alpha_0\oplus 1$, then Bob's guess is successful if $a\oplus b = 1$. Thus, the probability for success of Bob's guess in this case is,
\begin{eqnarray}
P_{II}\equiv P_{\beta = 1} &=&  \frac 12 (p(a\oplus b = 0|0,1)+p(a\oplus b = 1|1,1)). \label{PII}
\end{eqnarray}
It is not difficult to show that the CHSH parameter, (\ref{CHSH}), can be expressed as $S = 2(P_I+P_{II})$, or equivalently with the L1 norm as $M^* = P_I+P_{II}$. 

It is instructive to examine the PR box where $P_I = P_{II} = 1$ because PR correlations are deterministic and defined by $a\oplus b = xy$. Using a PR box Bob can always successfully guess the $\alpha_\beta$ bit of Alice. However, he can not guess both as that would imply signaling. The mutual information in this case is
\begin{eqnarray}
I &=& I(\alpha_0 : g|\beta=0) + I(\alpha_1 : g|\beta=1) = 2,
\end{eqnarray}
and thus $I>m$.

For isotropic boxes we have
$p(a\oplus b|x,y) = \frac 12 (1+\gamma)$, as described in figure \ref{fig:PolytopeIsotropic}.   
It was shown in \cite{pawlowski_information_2009}
that IC is violated for this protocol if 
\begin{eqnarray}
E_I^2 + E_{II}^2 >1,
\end{eqnarray}
 where $E_i = 2P_i-1$ . Notice this implies that IC is violated for all $\gamma>\frac 1{\sqrt 2}$, thus matching Tsirelson's bound.

\begin{figure}[htbp]
 
    \includegraphics{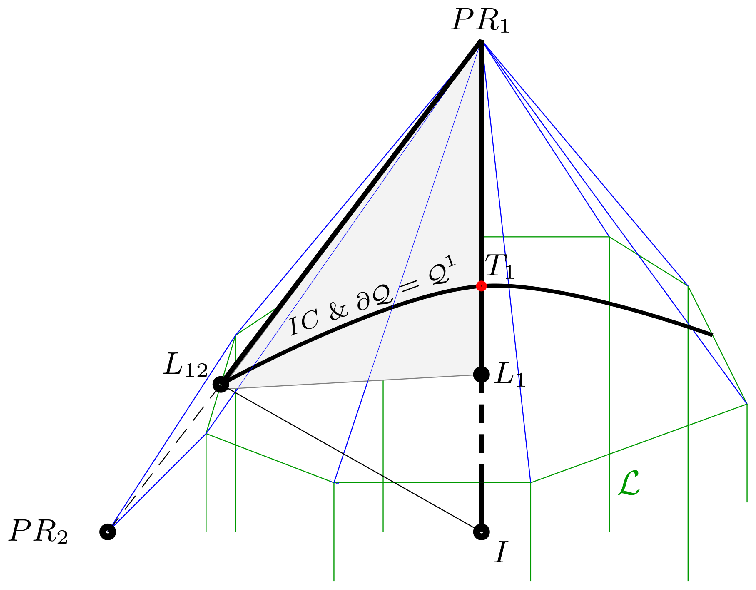}

    \caption{A slice of the polytope where IC defines the quantum set, where  		
$P^{12}_{\gamma,\beta} = \gamma PR_1+\beta L_{12} + (1-\gamma -\beta)I$ and
$L_{12} = \frac 12 (PR_1+PR_2)$. The box $L_{12}$ has a jpd consisting of 
elementary probabilities $p_{0000}=p_{0001}=p_{1110}=p_{1111} = \frac 14$.
All boxes within this slice have $\langle A_i\rangle = \langle B_i\rangle = 0, \forall i$, 
thus the TLM criterion (\ref{TLM})  is necessary and sufficient for being a quantum system.
Thus, for this portion of the polytope, IC, identifies the quantum set.}
    \label{fig:ICslice1}
 \end{figure}

\begin{figure}[htbp]

   \includegraphics{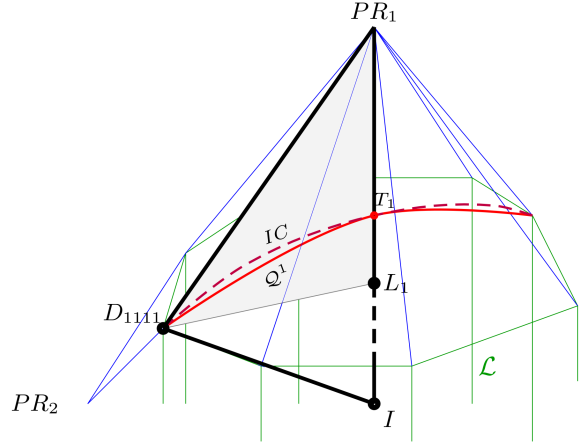} 
   
   \caption{The portion of the non-signaling polytope defined by $P^{1D}_{\alpha,\beta} = \alpha PR_1+\beta D_{1111} + (1-\alpha -\beta)I$,
   where $D_{1111}$ is the local box with a jpd consisting of $p_{1111} = 1$. In this slice of the polytope the boxes are unbiased, thus the TLM criterion
   does not give $\partial\mathcal{Q}$. This is a schematic representation and a plot of the curves are given in figure \ref{fig:LOslice2}. }     \label{fig:ICslice2}
\end{figure}

The early success of obtaining Tsirelson's bound via a physical principle, independent of the mathematical structure
of quantum theory, led to hope that IC would distinguish all physical systems from supra-quantal ones.
In 2009, Allcock et. al. \cite{allcock_recovering_2009}
examined various slices of the non-signaling polytope and found that for certain portions IC = $\partial\mathcal{Q}$, while for others IC is a necessary but not sufficient criterion to define quantum systems.
Specifically, they examined `noisy' PR boxes involving linear combinations of the noise box, PR, D, and L boxes.
\begin{eqnarray}
PR_{\gamma,\beta} &=& \gamma PR + \beta B + (1-\gamma - \beta)N. \label{noisyPR}
\end{eqnarray}
The various slices involve $B = D_{1111}$ (a deterministic box only outputting 1 for $a$ and $b$) and $B = L_{12} =  \frac 12 (PR+PR_i)$ (a box on $\partial\mathcal{L}$ halfway between two different PR boxes).
The results are schematically show in figures \ref{fig:ICslice1} and \ref{fig:ICslice2}. A detailed view of figure \ref{fig:LOslice2} is given later.

It must be stressed that the failure of IC to match $\partial\mathcal{Q}$ is not necessarily a failure of IC to single out physical systems. Results at this point in time rely on particular protocols to explore IC and it is possible that future protocols might single out $\partial\mathcal{Q}$.

\subsubsection*{Negative probability representation of IC}

The inequalities above using the van Dam protocol can all be recast in terms of a jqpd. The
success probabilities (\ref{PI}) and (\ref{PII}) can be expressed in terms of subsets of the jqpd,
\begin{eqnarray}
E_I &=& \sum_{a_0,a_1,b_0,b_1} (-1)^fp_{a_0a_1b_0b_1} \delta_{a_0=a_1}, \nonumber \\
E_{II} &=& \sum_{a_0,a_1,b_0,b_1} (-1)^fp_{a_0a_1b_0b_1} \delta_{a_0\neq a_1}.
\end{eqnarray}

For the isotropic boxes, this yields, $E_I = E_{II} = 4{(1+2\gamma)\over 16} - 4{(1-2\gamma)\over 16} = \gamma$
and the boundary $E_I^2+E_{II}^2 = 1$ when $\gamma = \frac 1{\sqrt 2}$.  The detailed boundaries in figure \ref{fig:LOslice2}
were generated using jqpds.

\section{Macroscopic Locality}

The principle of macroscopic locality (ML) stems from Bohr's correspondence principle: in the large $N$ limit, non-classical features should diminish and classical physics should be regained. As such, it can be phrased the following way:
\begin{quote}
\em As a microscopic system becomes macroscopic, we should regain classical correlations.
\end{quote}
In multipartite entangled systems, this corresponds to a system consisting of many identical non-local
systems (the microscopic systems) 
transforming into one describe with local statistics (the macroscopic system). The macroscopic variables behave as classical objects
and thus do not violate any Bell inequality. It is important that the macroscopic variables no longer contain
information about specific microscopic systems, thus an average is performed to obtain the macroscopic variable.

Specifically, we consider $N$ independent and identically distributed systems having a conditional probability distribution
$p(a^i,b^i|xy), i=1,\cdots , N$. The inputs, $x,y$ are fixed and each receive a beam of particles that are diverted to
the two detectors. Alice receives intensities $I_0, I_1$ and Bob receives $J_0,J_1$, the sum total of particles reaching each
detector. It is important to note that microscopic information is erased, i.e. there is no longer any information pairing microscopic outcomes.
Binary random variables for the macroscopic outcomes are defined as follows,
\begin{eqnarray}
\alpha_x = 0 & \;\; \mbox{if}\;\; & I_{0|x} - I_{1|x} \geq 0, \hskip 0.5 in  \beta_y = 0 \;\; \mbox{if} \;\;  J_{0|x} - J_{1|x} \geq 0   \nonumber \\
\alpha_x = 1 & \;\; \mbox{if}\;\;  & I_{0|x} - I_{1|x} < 0, \hskip 0.5 in  \beta_y = 1 \;\; \mbox{if} \;\;  J_{0|x} - J_{1|x} < 0.  \label{MLproc}
\end{eqnarray}

\begin{figure}[htbp]
	\centering
	   \includegraphics{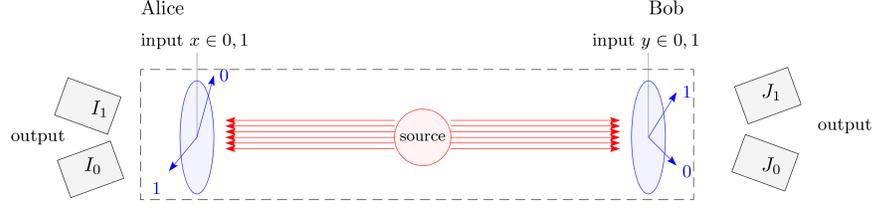}

   \caption{Macroscopic Locality considers $N$ copies of a non-signaling box where Alice and Bob lose the pair-wise identity of the microstate boxes. They record
		only the total number arriving at the two detectors for a fixed input. ML states that in the $N\rightarrow\infty$ limit, the statistics
		for physically realizable systems are local.}
   \label{fig:ML}
\end{figure}

The macroscopic locality principle (ML) states that physically allowed systems
admit a proper probability density for the intensities in the limit that $N\rightarrow \infty$ and under 
the assumption that the measurement devices cannot detect fluctuations less than
order $\sqrt N$. In terms of the random variables $\alpha_x, \beta_y$ the statistics should
satisfy the CHSH inequality in the macroscopic limit.

To investigate the effectiveness of ML we begin by examining PR correlations. It is straightforward
to show that any number of PR boxes maintain the PR correlations. As PR correlations are $\pm 1$,
the number of  outcomes of 0's and 1's are identical for Alice and Bob with inputs $(x,y) = (0,0), (0,1), (1,0)$
and thus the macroscopic correlation in these cases remains  $+1$. Under inputs $(1,1)$ the perfect anti-correlation
leads to perfect anti-correlation of the macroscopic random variables. Thus, regardless of the number
of copies, PR boxes map into PR boxes in the macroscopic limit, thus violating the principle.

In \cite{navascues_glance_2010} 
it was proved that systems satisfying ML are equivalent to the NPA hierarchy $Q^1$. Thus the Tsirelson bound is
obtained by ML. However, as it is known that $Q\subset Q^1$, ML does not eliminate all supra-quantal systems.
It is also possible to identify a system that satisfies ML but violates IC, suggesting that IC is a stronger constraint.
However, within the van Dam protocol there are boxes which violate ML yet satisfy IC. Again, it should be stressed that
future protocols might place stronger constraints on physically allowable states as defined by IC.

\subsection{Negative probability analysis of ML} 

As ML is a principle pertaining to the statistics of microscopic and macroscopic systems (and not of particular communication
protocols) it is straightforward to represent macroscopic systems with jqpds. For simplicity we examine isotropic boxes
and the macroscopic systems explicitly for odd $N= 1$ to 9. As the microscopic elementary probabilities take the form
$p_{a_0a_1b_0b_1} = {1+ (-1)^f2\gamma\over 16}$, the elementary probabilities for the combined system take the form,
\begin{eqnarray}
p_{a_0^1a_1^1b_0^1b_1^1\cdots a_0^Na_1^Nb_0^Nb_1^N} &=& \frac 1{16^N}\prod_{i = 1}^N (1+(-1)^{f^i}2\gamma). \label{MLisotropic}
\end{eqnarray}
Local processing then allows one to construct the marginal probabilities $p(\alpha_x,\beta_y)$ for the macroscopic binary variables.

\begin{figure}[htbp]
   \centering
   \includegraphics[width= 4.5 in]{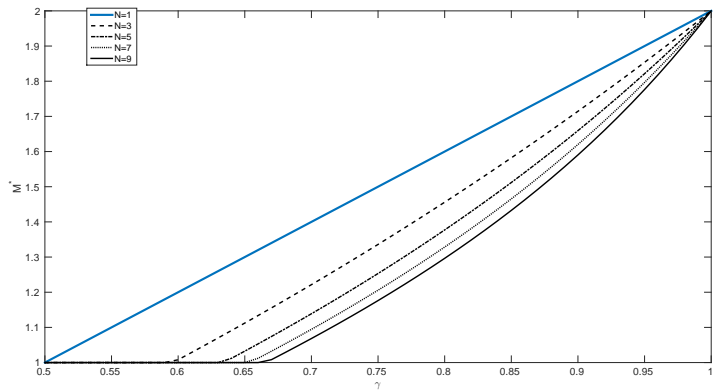} 
   \caption{The L1 norm, $M^*$, for $N$ copies of the isotropic box (\ref{MLisotropic}) processed into  binary macroscopic variables (\ref{MLproc}).
   Only odd terms are displayed as cases with $N$ even have an asymmetry due to the $I_i, J_j = 0$ term which  becomes negligible in the large $N$ limit.}
   \label{fig:ML_NP}
\end{figure}

\begin{figure}[htbp]
   \centering
   \includegraphics[width=\textwidth]{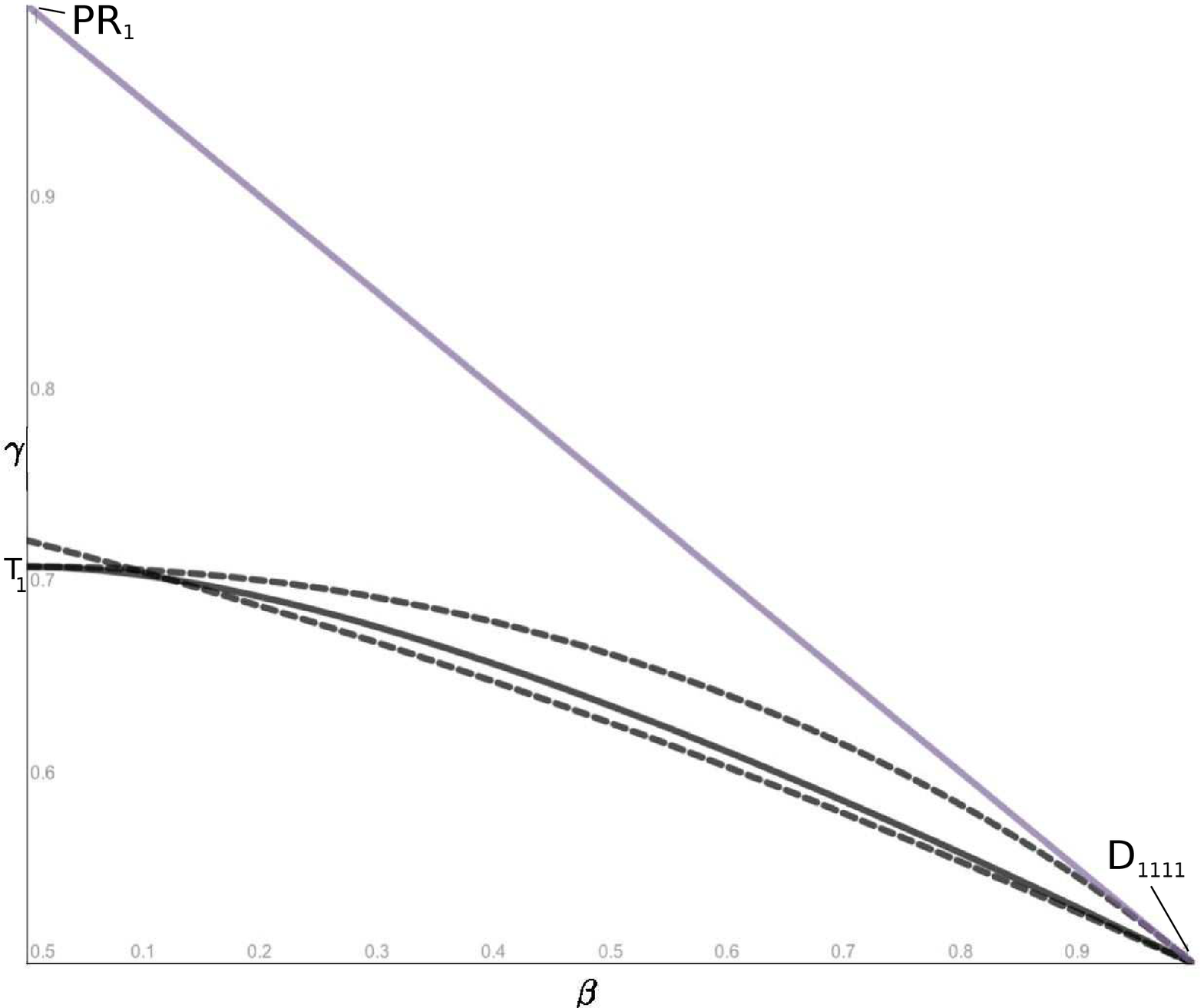} 
   \caption{The 10 term $LO^2$ inequality as compared to NS, IC, and $\mathcal{Q}^1$. The upper solid line is the edge of the NS polytope. The lower
   solid line is $\partial\mathcal{Q}^1$. The upper dashed line is the upper bound of IC-satisfying boxes while the lower dashed curve is the upper
   bound of the $LO^2$ satisfying set of boxes. This slice of the polytope is a detailed portion of that shown in figure \ref{fig:ICslice2}.
   Note well that, while IC does reach the Tsirelson bound, there are many more supra-quantal systems satisfying IC as compared to $LO$. However,
   $LO$ does not reach the Tsirelson bound. Recall also that ML $= \mathcal{Q}^1$ and NTCC is violated for systems $M^* > 1.508$, or  $\gamma > 0.754$.}
   \label{fig:LOslice2}
\end{figure}

In Figure \ref{fig:ML_NP}
the L1 norm is plotted for isotropic boxes from $L_1$ to $PR_1$. It is seen that as the number of microscopic boxes is increased, more states within $\mathcal{Q}$ become describable by a proper probability distribution, and convergence towards $\gamma =\frac 1{\sqrt 2}$ is suggested.

\section{Local Orthogonality (LO) / Exclusivity (E)}

In 1960 Specker (see \cite{cabello_speckers_2012})
conjectured a principle potentially defining $\mathcal{Q}$: a set of pairwise decidable propositions is jointly decidable.
This principle may not have as much physical justification as the previous examples, however it can be used
to limit the range of correlations. In 2012 Fritz et. al. \cite{fritz_local_2013} proposed a related principle, local orthogonality, which
suggests that events in a probability space involving different outcomes for the same context, labeled {\em orthogonal}, or {\em exclusive},
events, have probabilities respecting standard probability. In other words, 
\begin{quote}
\em Events which are pairwise exclusive are jointly exclusive. 
\end{quote}
Explicitly, the sum of probabilities involving mutually
orthogonal events must sum to unity or less. 

The Exclusivity principle (E), also know as Consistent Exclusivity, applies similar reasoning to scenarios involving contextuality, but not 
necessarily non-locality. As this falls outside of the non-locality discussion here, we refer the reader to Cabello's introduction of the principle \cite{cabello_simple_2013}

Considering a 2222 box, it was proved in \cite{fritz_local_2013}
that LO is equivalent to NS. The LO inequality in this case takes the following form,
\begin{eqnarray}
\sum_{a\neq a_0,b = 0}^1 P(ab|xy') + \sum_{b=0}^1 P(a_0b|xy) &\leq& 1, \nonumber \\
P(10|01)+P(11|01) + P(00|00)+P(01|00)  &\leq &1  \label{LO1}
\end{eqnarray}
where in the second line we have set $a_0 = 0, x = 0$. The proof is straightforward, however is trivial
when expressed in terms of a jqpd. 
As was shown in \cite{oas_exploring_2014}  
the non-signaling condition is equivalent to the existence of a jqpd. 
In addition, since all observable marginal probabilities are
proper (real and nonnegative), any sum of observable conditional probabilities will necessarily be
less than or equal to 1. Thus non-signaling implies LO. To show that LO implies NS, we only
need to consider the inequality  (\ref{LO1}) and express it in terms of elementary probabilities,
\begin{eqnarray}
p_{1\cdot \cdot 0}+ p_{1\cdot \cdot 1}+p_{0\cdot 0\cdot}+ p_{0\cdot 1\cdot} &\leq & 1. 
\end{eqnarray}
Where $\cdot$ implies a sum over that random variable's outputs. Thus we have,
\begin{eqnarray}
p_{1\cdot\cdot\cdot} + p_{0\cdot\cdot\cdot} &=& p_{\cdot\cdot\cdot\cdot} = 1.
\end{eqnarray}
Thus, satisfying this single LO inequality is sufficient to show that a jqpd exists and thus
NS is satisfied. Extending this proof to bipartite systems beyond binary inputs and outputs
is straightforward.

In order to tighten the bound on systems copies of systems are made, leading to a hierarchy of levels
to which the principle may be applied. One only need to go to a single copy, $N=2$, to show a violation
for PR boxes. For $LO^2$ the conditional probabilities are of the form $p(aba'b'|xyx'y') = p(ab|xy)p(a'b'|x'y')$,
and it is easy to verify that the following inequality is comprised of probabilities for events which
are orthogonal,
\begin{eqnarray}
p(0000|0000) + p(1110|0011) + p(0011|0110) \nonumber \\
+ p(1101|1011)+ p(0111|1101) \leq  1. \label{LO2_5term}
\end{eqnarray}
Recalling the PR probabilities (\ref{PRprob}), we see that each term is equal to $\frac 14$ and thus the inequality is violated.
 
 The inequality (\ref{LO2_5term}) is inadequate to analyze systems below the PR box and a larger inequality
 involving 10 probabilities is shown to be maximal (modulo the trivial inequalities composed of those with single set of inputs),
 \begin{eqnarray}
 p(1111|0000) +p(1100|1010) +p(0100|1100) +p(0011|0001) \nonumber \\
  +p(0010|0111) + p(1011|0000) + p(0101|1100) + p(1101|1100) \nonumber \\
  + p(1010|0110) + p(1001|0100)  \leq 1. \label{LO10}
 \end{eqnarray}
 One can verify that all events in (\ref{LO10}) are pairwise orthogonal. Examining the set of isotropic boxes
 the maximal box satisfying $LO^2$ yields $\gamma \equiv 0.7208$ and thus LO, at this level, does not reach the Tsirelson bound of $\gamma = \frac 1{\sqrt 2} = 0.707$.
 
 \subsection{LO as cast within NP}

As orthogonal (or exclusive) events involve different outcomes of the same measurement, it is obvious that such events
can not share any of the same probabilities of elementary events, $p_{a_0a_1b_0b_1}$. Thus, the LO principle can be recast as,

\begin{quote}
LO: Any set of observable marginal probabilities that do not share any elementary probability, $p\in$ jqpd, must sum to unity or less.
\end{quote}
 
It is clear that all classical systems, admitting a jpd, satisfy the LO principle as proper probability is monotonic. 
In addition, for contextual systems it is clear that this principle is not trivial (as negative probability is non-monotonic) as will be demonstrated shortly.
Lastly, if ML is satisfied as $N=\infty$ clearly LO is satisfied as there exists a jpd.

 \subsubsection{Status of LO and E}
 
 The task of finding maximal, non-trivial, inequalities is a proven hard task. In order to construct inequalities, even at the $N=2$ level,
 graph theoretical techniques were employed \cite{sainz_exploring_2014}, and later in \cite{acin_combinatorial_2015} (AFLS). It was
 found that not only does $LO^\infty$ not reach $\mathcal{Q}$ (nor $\mathcal{Q}^{1+AB}$), but the infinite level of the consistent
 exclusivity hierarchy, $\mathcal{CE}^\infty$(of which $LO$ is a special case) is not even a convex set. Thus linear combinations
 of systems satisfying $LO^\infty$ may not satisfy $LO^\infty$, in the parlance of the field $LO$ systems can be `activated.'
 In \cite{acin_combinatorial_2015} {\em extended consistent exclusivity}, $\widetilde{\mathcal{CE}}$ was introduced\footnote{The basic idea is to only 
 include those $\mathcal{CE}$ systems, $p$, for which a linear combination with a system within $\mathcal{Q}$, $q$ resulted in a system
 that satisfied $\mathcal{CE}$, i.e. are not activated.} and shown to be equivalent to $\mathcal{Q}^1$.

The exclusivity principle is still being utilized to explore many contextuality scenarios (contextuality without direct influence)
Recently Cabello \cite{cabello_exclusivity_2014} has claimed to demonstrate that systems above the Tsirelson bound violate
E for even two copies. However, it must be emphasized that this result only holds under the strong assumption of ``sharp measurements,"
non-demolition measurements that minimally disturb the state. Thus, this result can not be suggested to solve the problem
as entangled photonic systems involve demolition measurements, hence the fundamental question still remains.

\section{Final Remarks and Conclusions}
Summarizing the state of affairs with the principles reviewed, we begin by noting that NS opened up the
range of potentially physically possible states constrained only by relativity. Exploiting non-signaling systems
in computation and communication complexity scenarios, it was found that extreme non-signaling systems, PR boxes,
could lead to the unsavory effect of being able to make any communication task trivial, independent of the size
of the task. This also leads to extraordinary computational powers not expected in the physical world. 
NTCC imposes the constraint that such trivial communication should not be a feature of our universe and
those systems violating are then unphysical. In constraining the range of boxes, NTCC has only eliminated
those above $S>3.266$, and a small region near the boundary of the $\mathcal{NS}$ polytope via
nonlocality distillation \cite{brunner_nonlocality_2009}. It is hoped that better distillation methods might 
tighten the bound between NTCC systems and the quantum set. This hope has been dashed as very recent work
have introduced a supra-quantal set that satisfies NTCC (discussed in the next section). 

Information causality received much attention when introduced, as it was the first physical principle to reach
the Tsirelson bound. It is still the one principle that holds hope in achieving the goal of demarcating $\mathcal{Q}$,
with the introduction of better protocols. However such protocols are slow to come by. It has been shown that
certain trajectories do yield $\partial\mathcal{Q}$, while the canonical slice (linear combination of the $PR - I - D$ boxes)
IC fares worse than other principles (ML, LO).

Macroscopic Locality appears as a very natural physical principle. In the large $N$ limit, we do not expect to observe macroscopic
nonlocality. However, in the introductory paper \cite{navascues_glance_2010}, it was proved that the set of systems satisfying ML is equivalent
to the first level of the NPA hierarchy, $\mathcal{Q}^1$. Thus, ML, while eliminating a large amount of supra-quantal systems, does not
characterize the quantum set.\footnote{Recently, an extended principle {\em Macroscopic Non-Contextuality} (MNC) has been proposed
which applies the macroscopic limits to contextual systems not involving entanglement \cite{henson_macroscopic_2015}. 
By employing the graph theoretical techniques of AFLS \cite{acin_combinatorial_2015}, Henson and Sainz claim to improve the
bound for MNC to $\mathcal{Q}^{1+AB}$ However, this result has not yet been published.}

Local Orthogonality has greatly limited the number of supra-quantal states yet has not achieved the significant milestone of
of reproducing the Tsirelson bound. As mentioned above, the graph theoretical analysis of AFLS \cite{acin_combinatorial_2015} proved that LO does not match $\mathcal{Q}$ for any
level of the hierarchy. However, examining orthogonal events, or, more accurately, exclusive events (the E principle) has been fruitful in examining 
contextuality scenarios not in the non-local domain.

\subsection{Almost Quantum theory}

As the five principles have not demonstrated the ability to constrain to the quantum set, some have suggested
that perhaps quantum theory needs to be altered slightly. 
Based on the convergence of entirely different approaches
to the problem, it has recently been proposed by Navascues et. al. \cite{navascues_almost_2015} 
that a slight change to quantum theory might be warranted.
Specifically, if one weakens the second constraint on the projectors presented in the NPA section as follows
\begin{equation}
[E_x^a,E_y^b] = 0 \;\;\;\;  \longrightarrow \;\;\;\;  [E_x^a,E_y^b]|\psi\rangle = 0,
\end{equation}
one obtains the $\mathcal{Q}^{1+AB}$ set of the NPA hierarchy\footnote{Note that this imposes state-dependent conditions on the projectors so is thus a weakening of standard quantum theory.}.
It is suggested that this modification, termed {\em Almost Quantum} theory, or $\tilde{\mathcal{Q}}$, and characterized by 
the $\mathcal{Q}^{1+AB}$ level of the NPA hierarchy, might be appropriate theory to describe physical systems.
First of all, it has been shown that all of the principles discussed, except for IC which is still undecided,  
are satisfied by $\tilde{\mathcal{Q}}$ systems. This implies that these principles do not suffice to characterize the quantum set. Secondly, approaches
removed from quantum information have been shown to yield almost quantum systems. 

Specifically, in the attempt to formulate a theory of quantum gravity
one approach has been to use the histories formulation to incorporate a spacetime character into quantum theory (this applies to the causal set approach
to quantizing spacetime). Along these lines Sorkin introduced quantum measure theory \cite{sorkin_quantum_1994}
 and proposed that the lack of third order interference among histories (e.g. in a multi-slit interference experiment the result is characterized by pairwise
 interference terms and no higher order terms). Recent work by Dowker et. al. \cite{dowker_histories_2014} have 
 found that within this approach the $\mathcal{Q}^{1+AB}$ systems are singled out.
Further evidence pointing towards Almost Quantum theory comes from certifiable randomness scenarios. In \cite{de_la_torre_maximally_2015}, it has been
found that in a tripartite scenario the set $\mathcal{Q}^{1+ABC}$, an analog to the almost quantum set, can certify maximal randomness.

An argument in favor of Almost Quantum theory is that it would make the problem of identifying almost quantum systems decidable and efficient, while not violating any of the natural principles introduced.
That is, the question as to whether  $p(a_x,b_y)\in \tilde{\mathcal{Q}}$ is efficiently solved, while answering the question as to whether  
$p(a_x,b_y)\in \mathcal{Q}$ is known to be hard, and possibly undecidable.
There is still much exploration to be done and a physical theory of Almost Quantum mechanics is still lacking, thus it is an open question whether
our universe obeys almost quantum, or strictly quantum mechanics.

The main results of these attempts is summarized in table below.
\begin{table}[ht]
   \centering
   \begin{tabular}{cccccc}
{\small Principle} & {\small PR eliminated?} & {\small Tsirelson bound?} & { $\mathcal{Q}^1$} & $\mathcal{Q}^{1+AB}$ & $\mathcal{Q}$ \\\hline
NS & no & no & $>$ & $>$ &  $>$ \\
NTCC & yes & no &  $>$ &  $>$ &  $>$ \\
IC & yes & yes & $>,=$? & $>$? & $>$? \\
ML & yes & yes & = & $>$ &  $>$ \\
LO & yes & no & $<,>$ & $>$ &  $>$ \\
$\tilde{\mathcal{Q}}$ & yes & yes & $<$ & $=$ &  $>$ \\
\end{tabular}
    \caption{Summary of principles reviewed in this discussion. PR boxes violate all principles except NS. Those principles that admit systems above Tsirelson bound are labeled `no,'
    and those that admit systems above the levels of the NPA hierarchy are indicated by $>$
    are IC and ML. Within the NPA hierarchy, those principles that do not reach those levels are indicated with $>$, those that match by $=$, and those that offer stronger
    bounds by $<$. Note that LO is tighter than $\mathcal{Q}^1$ in places and weaker in others, while IC's scope has not been formally determined at this time (indicated by a question mark).}
   \label{tab:booktabs}
\end{table}
Those principles that admit systems above the Tsirelson bound are labeled `no,'
    and those that admit systems above the levels of the NPA hierarchy are indicated by $>$. Note that LO is tighter than $\mathcal{Q}^1$ in places and weaker in others, while IC's scope has not been formally determined at this time (indicated by a question mark).

The search continues for physical or informational principles that define quantum theory. It is surprising that the reasonable conditions proposed:
\begin{itemize}
\item[] can not violate relativity;
\item[] communication and computation should be trivial;
\item[] should not be able to receive more information than is sent; and
\item[] should not be able to observe non-classical effects at the macroscopic scale  
\end{itemize}
are satisfied by systems that are non-physical. It is possible that some obscure, seemingly unphysical, constraint might
single out $\mathcal{Q}$, or it may be that we learn something new about nature if it is ever defined.

It is clear that novel views and tools could assist with this goal. Tools such as Generalized Probability Theory (GPT), where one
defines a notion of states and evolution, puts the local, quantum, and non-signaling systems on the same footing, to help highlight differences (see \cite{janotta_generalized_2014} for a review).

Here we have utilize a theory of negative probability to investigate these principles. It should come as no surprise that jqpds can describe the range of systems as it is designed to yield observed marginal probabilities and the standard analysis is often cast in terms of the latter. One motivation 
of this approach is to keep  as close as possible to standard probability theory allowing use of its rich toolset. 
Earlier work \cite{oas_exploring_2014}
demonstrated how negative probability can characterize non-local systems and can be related to other approaches
characterizing contextuality \cite{de_barros_unifying_2014}.

The value of this approach is to place
these principles within the same framework, while adhering to the device-independent approach. For example, one sees similarities between the local orthogonality principle, $LO^\infty$, and macroscopic locality. The jqpd is identical for these two cases, yet the conditions imposed (proper jpd for ML and subsets of probabilities for orthogonal events for LO) show that ML is the stronger condition in this case -satisfaction of ML guarantees satisfaction of LO, but not necessarily the reverse. 

It is possible that examining non-local systems in a new light may provide guidance in the search for defining principles.

\paragraph{Acknowledgements.}
The authors wish to thank Ehtibar Dzhafarov for allowing us to participate in the 2014 Winer Memorial lectures. We would also like to thank the other participants
allowing the enlightening and spirited discussions, among them Samson Abramsky, Andrei Khrennikov,  Janne Kujala, Jerome Busemeyer, Guido Bacciagaluppi, Arkady Plotnitsky, and
Louis Narens. Lastly, we like to thank Patrick Suppes, Claudio Carvalhaes, and Stephan Hartman for related discussions over the years.

\bibliographystyle{plain}
\bibliography{QuantumNP}

\end{document}